\documentclass[prc,twocolumn,amsmath,amssymb,floatfix,superscriptaddress]{revtex4} 

\usepackage{amssymb} 
\usepackage{graphicx}
\usepackage{dcolumn}
\usepackage{bm} 
\usepackage{epsfig}
\usepackage[usenames]{color} 
\definecolor{green}{rgb}{0.0,0.65,0.55}

\newcommand{ \rts}{\ensuremath{\sqrt{s_{_{\rm NN}}}}}

\newcommand{\detaabs}{$|\Delta\eta|$ }
\newcommand{\pt}{\ensuremath{p_t} }
\newcommand{\ptno}{\ensuremath{p_t}}
\newcommand{\deta}{\ensuremath{\Delta\eta} }
\newcommand{\dphi}{\ensuremath{\Delta\phi} }
\newcommand{\dphiabs}{\ensuremath{|\Delta\phi|} }
\newcommand{\detano}{\ensuremath{\Delta\eta}}
\newcommand{\dphino}{\ensuremath{\Delta\phi}}
\newcommand{\pttrig}{\ensuremath{p_{t}^{trig}} }
\newcommand{\ptass}{\ensuremath{p_{t}^{assoc}} }
\newcommand{\pttrigno}{\ensuremath{p_{t}^{trig}}}
\newcommand{\ptassno}{\ensuremath{p_{t}^{assoc}}}

\newcommand{\GeVc}{\ensuremath{\mathrm{GeV}/c}}

\newcommand{\dsqnwitharg}{\frac{d^2N}{d\Delta\phi\, d\Delta\eta}(\Delta\phi,\Delta\eta)}

\newcommand{\bEta}{b_{\Delta\eta}}
\newcommand{\bPhi}{b_{\Delta\phi}}

\newcommand{\Yjeta}{Y^{\Delta\eta}_{J}}

\newcommand{\dNdPhi}{\frac{dN}{d\Delta\phi}}
\newcommand{\dNdEta}{\frac{dN}{d\Delta\eta}}

\def\Deta{\mbox{$\Delta\eta$}}
\def\Dphi{\mbox{$\Delta\phi$}}

\def\Argab{\mbox{$\left[a,b\right]$}}
\def\Argpseven{\mbox{$\left[-0.7,0.7\right]$}}

\def\YDetadelta{\mbox{$Y_{slice}\left(\Deta;\delta\right)$}}
\def\YDetadeltathree{\mbox{$Y_{slice}\left(\Deta;\delta=0.3\right)$}}

\def\Yridge{\mbox{$Y_{ridge}$}}

\def\vtwo{\mbox{$v_{2}$}}

\begin{document}

\title{Long range rapidity correlations and jet production in high energy nuclear collisions}

\affiliation{Argonne National Laboratory, Argonne, Illinois 60439, USA}
\affiliation{University of Birmingham, Birmingham, United Kingdom}
\affiliation{Brookhaven National Laboratory, Upton, New York 11973, USA}
\affiliation{University of California, Berkeley, California 94720, USA}
\affiliation{University of California, Davis, California 95616, USA}
\affiliation{University of California, Los Angeles, California 90095, USA}
\affiliation{Universidade Estadual de Campinas, Sao Paulo, Brazil}
\affiliation{University of Illinois at Chicago, Chicago, Illinois 60607, USA}
\affiliation{Creighton University, Omaha, Nebraska 68178, USA}
\affiliation{Czech Technical University in Prague, FNSPE, Prague, 115 19, Czech Republic}
\affiliation{Nuclear Physics Institute AS CR, 250 68 \v{R}e\v{z}/Prague, Czech Republic}
\affiliation{Institute of Physics, Bhubaneswar 751005, India}
\affiliation{Indian Institute of Technology, Mumbai, India}
\affiliation{Indiana University, Bloomington, Indiana 47408, USA}
\affiliation{Institut de Recherches Subatomiques, Strasbourg, France}
\affiliation{University of Jammu, Jammu 180001, India}
\affiliation{Joint Institute for Nuclear Research, Dubna, 141 980, Russia}
\affiliation{Kent State University, Kent, Ohio 44242, USA}
\affiliation{University of Kentucky, Lexington, Kentucky, 40506-0055, USA}
\affiliation{Institute of Modern Physics, Lanzhou, China}
\affiliation{Lawrence Berkeley National Laboratory, Berkeley, California 94720, USA}
\affiliation{Massachusetts Institute of Technology, Cambridge, MA 02139-4307, USA}
\affiliation{Max-Planck-Institut f\"ur Physik, Munich, Germany}
\affiliation{Michigan State University, East Lansing, Michigan 48824, USA}
\affiliation{Moscow Engineering Physics Institute, Moscow Russia}
\affiliation{City College of New York, New York City, New York 10031, USA}
\affiliation{NIKHEF and Utrecht University, Amsterdam, The Netherlands}
\affiliation{Ohio State University, Columbus, Ohio 43210, USA}
\affiliation{Old Dominion University, Norfolk, VA, 23529, USA}
\affiliation{Panjab University, Chandigarh 160014, India}
\affiliation{Pennsylvania State University, University Park, Pennsylvania 16802, USA}
\affiliation{Institute of High Energy Physics, Protvino, Russia}
\affiliation{Purdue University, West Lafayette, Indiana 47907, USA}
\affiliation{Pusan National University, Pusan, Republic of Korea}
\affiliation{University of Rajasthan, Jaipur 302004, India}
\affiliation{Rice University, Houston, Texas 77251, USA}
\affiliation{Universidade de Sao Paulo, Sao Paulo, Brazil}
\affiliation{University of Science \& Technology of China, Hefei 230026, China}
\affiliation{Shandong University, Jinan, Shandong 250100, China}
\affiliation{Shanghai Institute of Applied Physics, Shanghai 201800, China}
\affiliation{SUBATECH, Nantes, France}
\affiliation{Texas A\&M University, College Station, Texas 77843, USA}
\affiliation{University of Texas, Austin, Texas 78712, USA}
\affiliation{Tsinghua University, Beijing 100084, China}
\affiliation{United States Naval Academy, Annapolis, MD 21402, USA}
\affiliation{Valparaiso University, Valparaiso, Indiana 46383, USA}
\affiliation{Variable Energy Cyclotron Centre, Kolkata 700064, India}
\affiliation{Warsaw University of Technology, Warsaw, Poland}
\affiliation{University of Washington, Seattle, Washington 98195, USA}
\affiliation{Wayne State University, Detroit, Michigan 48201, USA}
\affiliation{Institute of Particle Physics, CCNU (HZNU), Wuhan 430079, China}
\affiliation{Yale University, New Haven, Connecticut 06520, USA}
\affiliation{University of Zagreb, Zagreb, HR-10002, Croatia}

\author{B.~I.~Abelev}\affiliation{University of Illinois at Chicago, Chicago, Illinois 60607, USA}
\author{M.~M.~Aggarwal}\affiliation{Panjab University, Chandigarh 160014, India}
\author{Z.~Ahammed}\affiliation{Variable Energy Cyclotron Centre, Kolkata 700064, India}
\author{A.~V.~Alakhverdyants}\affiliation{Joint Institute for Nuclear Research, Dubna, 141 980, Russia}
\author{B.~D.~Anderson}\affiliation{Kent State University, Kent, Ohio 44242, USA}
\author{D.~Arkhipkin}\affiliation{Brookhaven National Laboratory, Upton, New York 11973, USA}
\author{G.~S.~Averichev}\affiliation{Joint Institute for Nuclear Research, Dubna, 141 980, Russia}
\author{J.~Balewski}\affiliation{Massachusetts Institute of Technology, Cambridge, MA 02139-4307, USA}
\author{O.~Barannikova}\affiliation{University of Illinois at Chicago, Chicago, Illinois 60607, USA}
\author{L.~S.~Barnby}\affiliation{University of Birmingham, Birmingham, United Kingdom}
% ...

\collaboration{STAR Collaboration}\noaffiliation


\begin{abstract}
The STAR Collaboration at RHIC presents a systematic study of high
transverse momentum charged di-hadron correlations at small azimuthal
pair separation \dphino, in d+Au and central Au+Au collisions at
$\rts = 200$ GeV. Significant correlated yield for pairs with large longitudinal
separation \deta\ is observed in central Au+Au, in
contrast to d+Au collisions. The associated yield distribution in
\detano$\times$\dphi can be decomposed into a narrow
jet-like peak at small angular separation which has a similar shape to
that found in d+Au collisions, and a component which is narrow in \dphi and
\textcolor{black}{depends only weakly on} $\deta$, the ``ridge''. 
Using two systematically independent analyses,
\textcolor{black}{finite ridge yield} is found to persist for trigger
$\pt > 6$ \GeVc, indicating that it is correlated with jet
production. The transverse momentum spectrum of hadrons comprising the
ridge is found to be similar to that of bulk particle production in the measured range ($2 < \pt < 4\ \GeVc$).

%
%
\end{abstract}


\maketitle
\noindent
\section{Introduction}
Measurements of inclusive hadron suppression
\cite{star_pt,phenix_incl} and di-hadron azimuthal correlations
\cite{star1,star2,star_highpTcoor} in ultrarelativistic nuclear collisions 
have provided important insights into the properties of hot QCD matter
\cite{star_white,phenix_white,brahms_white,phobos_white}. In
particular, the high transverse momentum (high
\pt) suppression \cite{star_highpTcoor} and low \pt\ enhancement
\cite{star2} of the correlated yield of hadrons recoiling from a high
\pt\ particle (azimuthal pair separation \dphi$\sim$ $\pi$) suggest a
dramatic softening of jet fragmentation in dense matter, arising from
strong partonic energy loss.

Studies of near-side (small \dphino) di-hadron correlations
in events containing a ``trigger particle'' at high $\pt > 6$
\GeVc{} reveal a \emph{jet-like} correlation at small angular separation
(small pseudo-rapidity pair separation  \detano\ and small \dphino)
which is unmodified in central Au+Au collisions relative to d+Au
\cite{star_highpTcoor}, suggesting that the dominant production
mechanism is jet fragmentation outside the dense medium. 

At lower trigger momentum, significant near-side correlated yield has been observed at large pair separation \deta$\sim1$ \cite{star2}, while for un-triggered correlations, longitudinal broadening at small \deta is seen \cite{star_deta,star_LLL}. However, inclusive hadron production at moderate $p_t < 6$ GeV/$c$ in central Au+Au collisions differs significantly from that observed in more elementary collision systems \cite{star_LamK,star_piprot}, indicating that jet fragmentation may not be the dominant hadron production mechanism in the kinematic region of these studies. For example, the large baryon/meson ratio observed at intermediate \pt is generally attributed to hadron formation by coalescence of constituent quarks \cite{Fries:2003kq,Hwa:2002tu}, which might also affect the di-hadron correlation structure \cite{Hwa:2004sw}.

In this paper, we present new measurements using the STAR detector to
explore the near-side correlation structure in Au+Au and d+Au
collisions at $\sqrt{s_{NN}}=200$ GeV, with emphasis on the \deta{}
shape and high-\pt{} trigger particles. In central Au+Au collisions,
significant associated yield at large \deta is observed for all
\pttrigno, including $\pttrig > 6$ \GeVc\, where jet fragmentation is
expected to be the dominant particle production mechanism. At large
\pttrig\ the near side correlation structure can be factored into a
jet-like peak, with properties similar to correlations in p+p
collisions, and an elongated contribution which is approximately
independent of \detano, which we therefore call {\it the ridge}.

Based on the earlier measurements and preliminary versions of some of
the results in the present paper, several models have been proposed to
explain the observed broadening of the near-side distributions and the
occurrence of the ridge. Models based on radiative partonic energy
loss suggest that the ridge arises from the coupling of induced gluon
radiation to the longitudinal flow of bulk matter \cite{Armesto_flow},
or from the coupling of radiation to transverse chromo-magnetic fields
\cite{Romatschke,Majumder}. Other models attribute the ridge to the
effect of elastic scattering of the jet in the flowing medium
\cite{Wong}, to medium heating by a jet \cite{Hwa_flow}, to radial
flow of bulk matter in coincidence with a jet trigger bias due to
energy loss \cite{Voloshin:2003ud,Shuryak}, or to long-range rapidity
correlations arising from a Color Glass Condensate initial state
\cite{CGCRidge,CGCRidge2}.

In order to address some of the model expectations, we study not only
the shape of di-hadron correlations in $\deta$ and $\dphi$, but also
the \pt dependence of the correlated yield. The ridge yield at high
$\pt^{trig}$ is examined using two systematically independent
assessments of the background contribution of uncorrelated
tracks. Comparison is made to d+Au reference data, to quantify the
modification of jet fragmentation due to interactions in the hot
medium.


\begin{figure*}[t]
\includegraphics[width=0.49\textwidth]{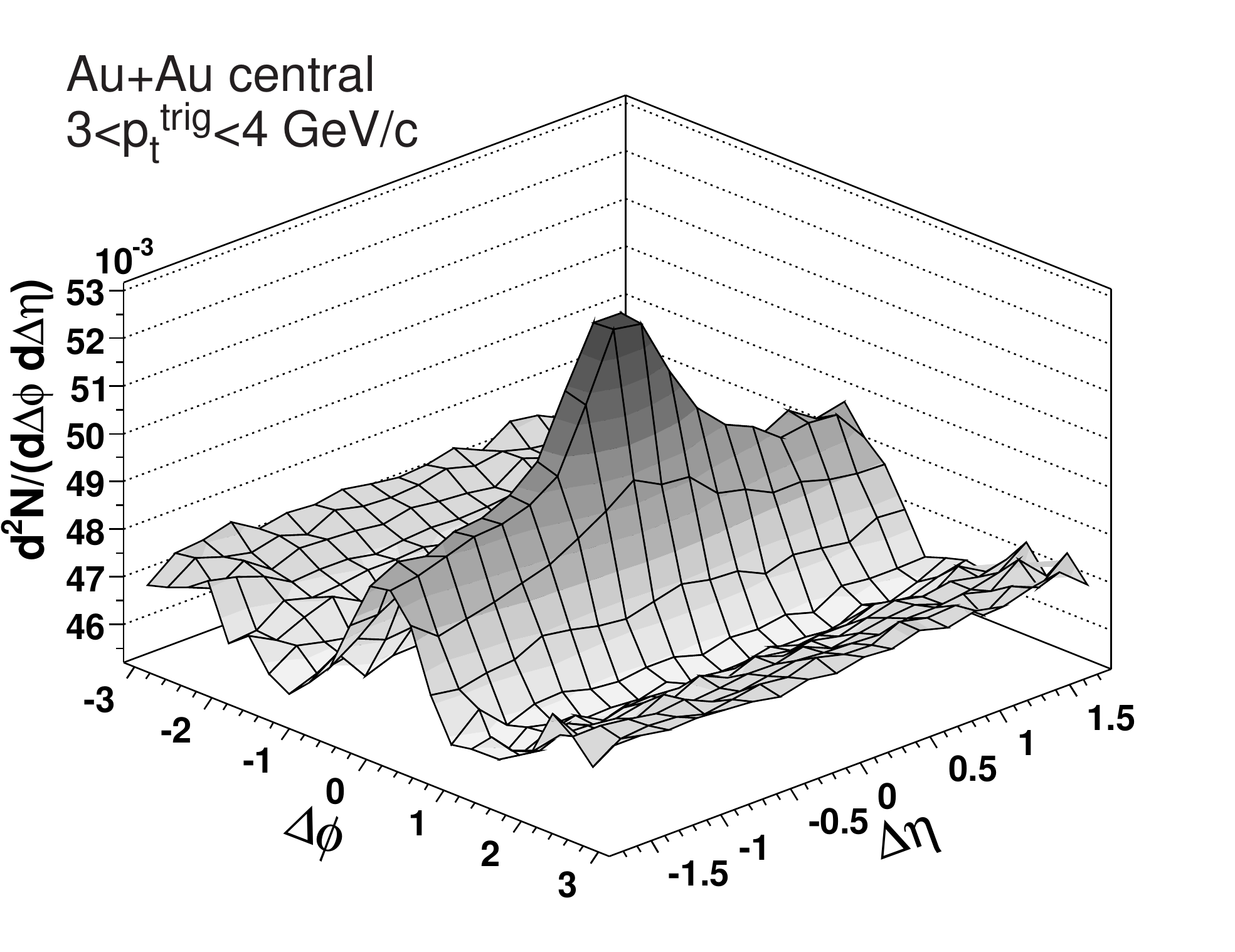}
\includegraphics[width=0.49\textwidth]{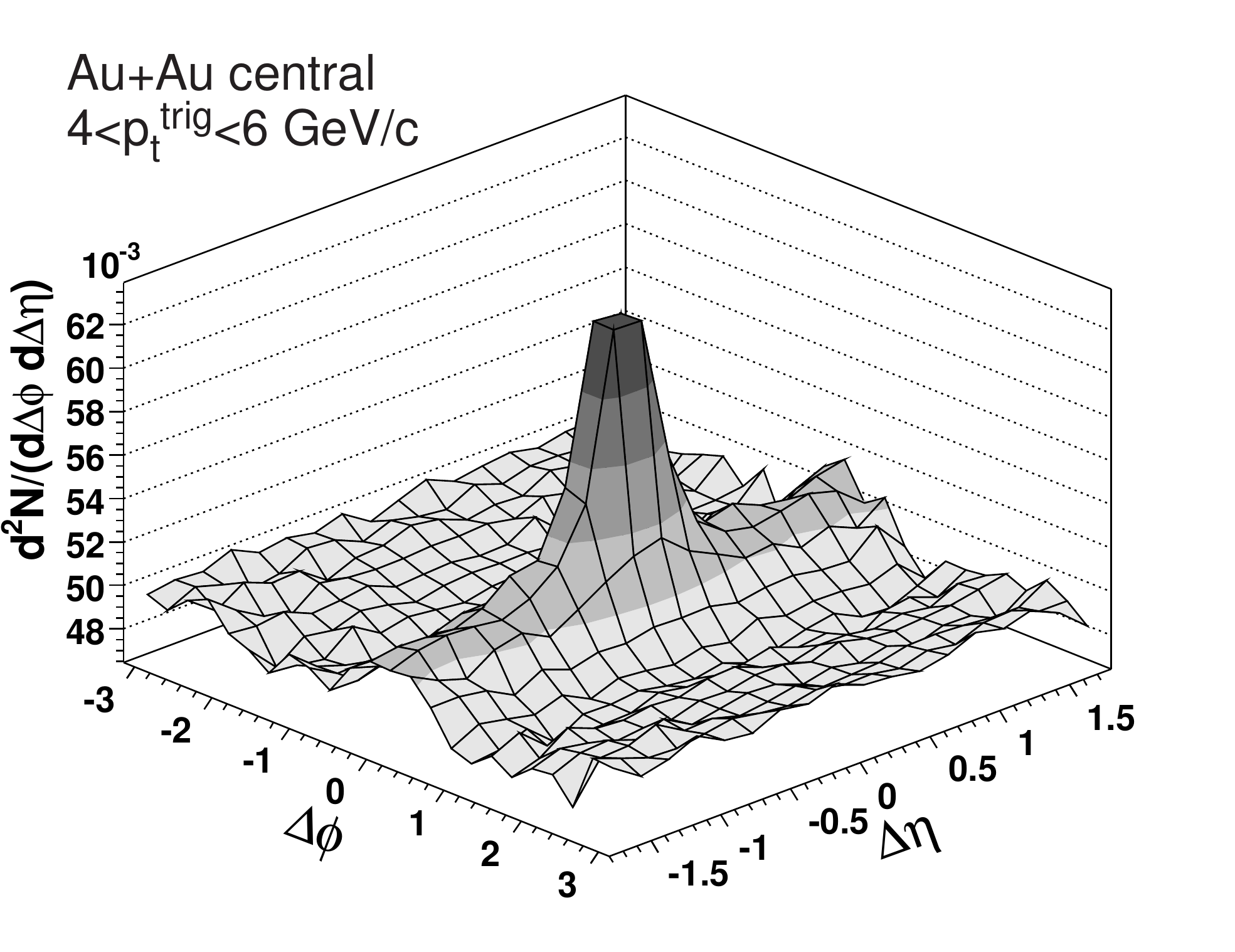}
\includegraphics[width=0.49\textwidth]{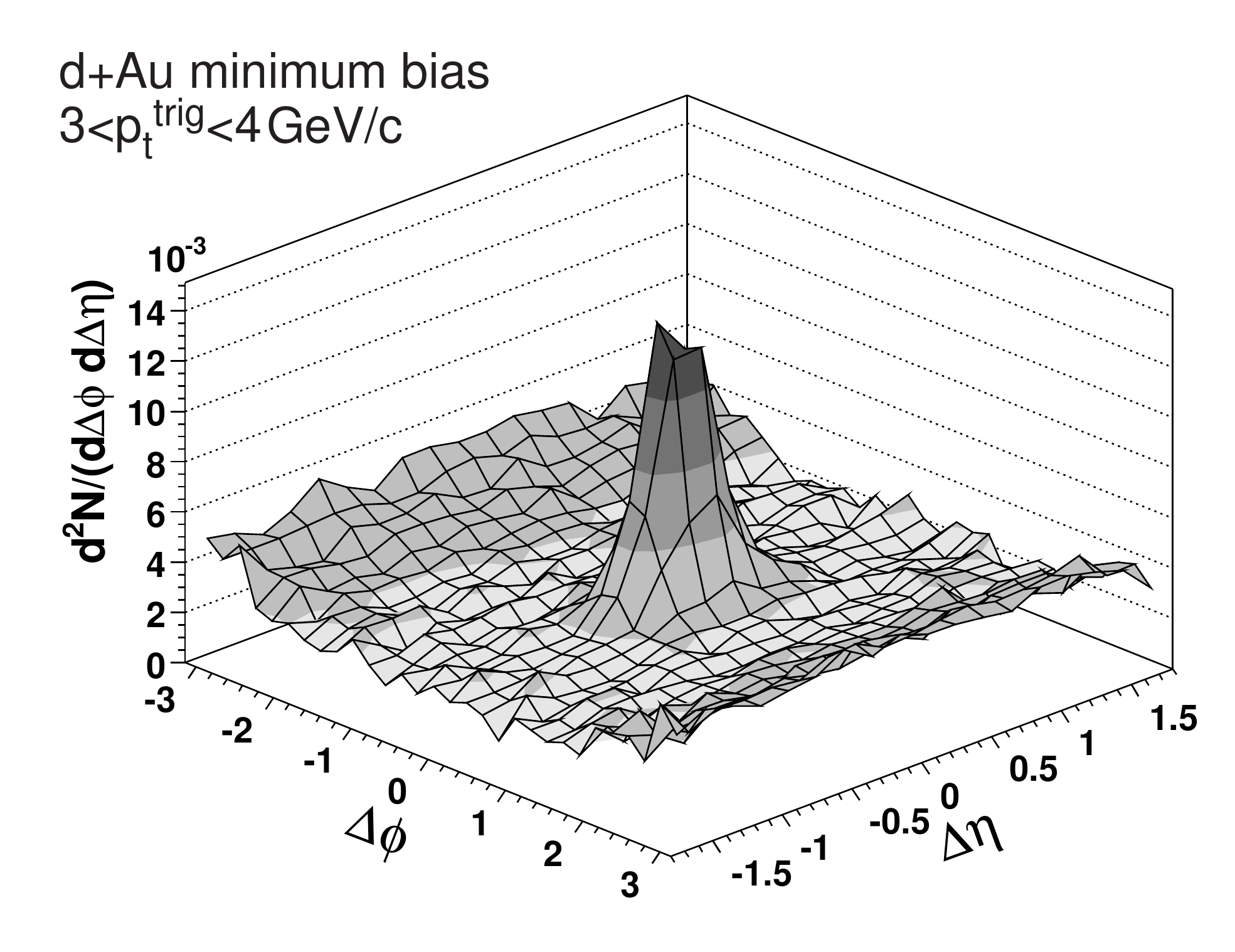}
\includegraphics[width=0.49\textwidth]{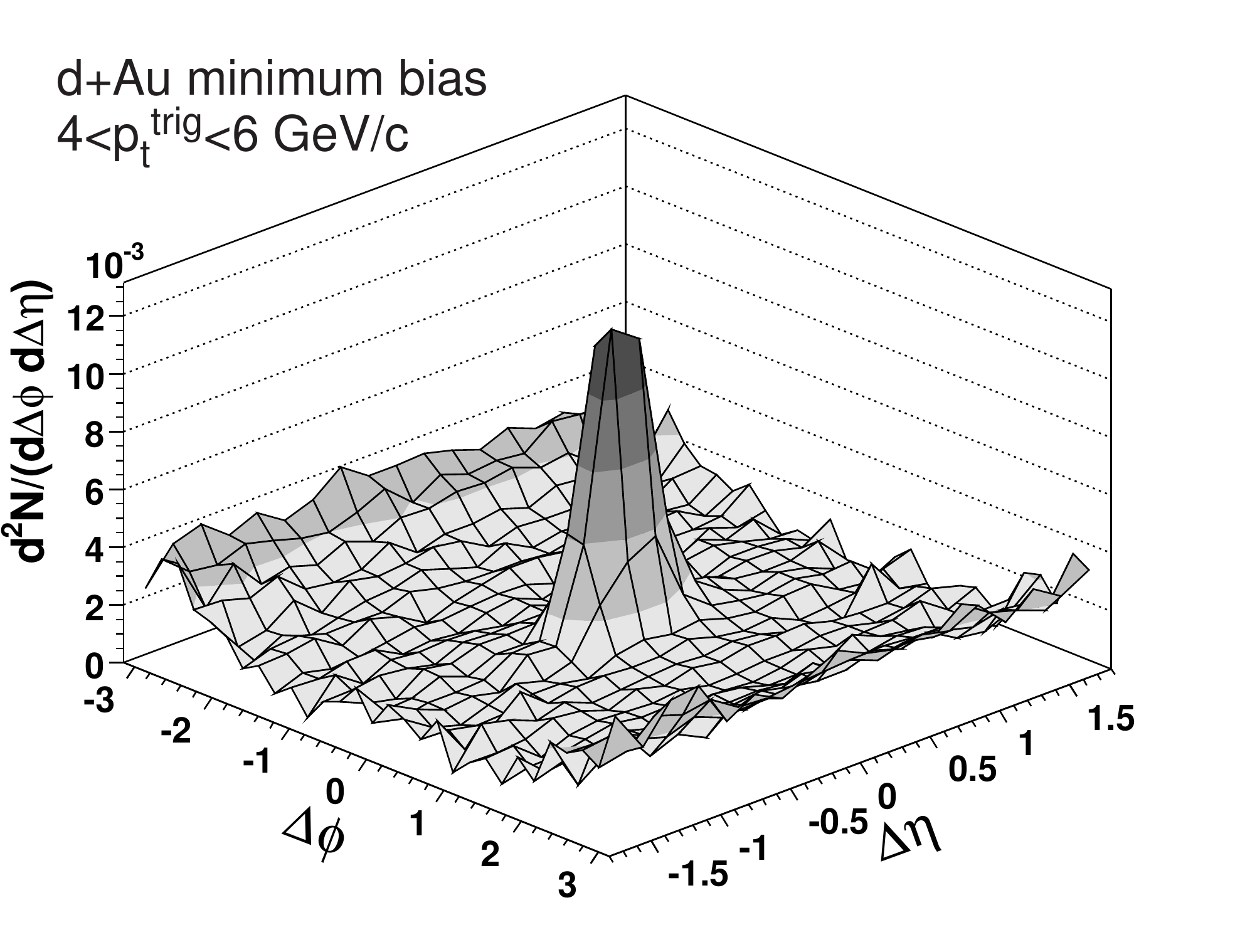}
\caption{\label{pr}
Charged di-hadron distribution (Eq.~\ref{eq:di-hadronDist}) for
$2\ \GeVc < \ptassno < \pttrigno$. Upper left: central Au+Au, $3 < \pttrigno < 4$
\GeVc; Upper right: central Au+Au, $4 < \pttrigno < 6$ \GeVc; Lower
left: minimum bias d+Au, $3 < \pttrigno < 4$ \GeVc\ ; 
Lower right: minimum bias d+Au, $4 < \pttrigno < 6$
\GeVc. Note different vertical scales. 
}
\end{figure*}


\section{Experimental setup}
The measurements were carried out by the STAR Collaboration at RHIC
\cite{star_detector}, using 12 million minimum bias
d+Au events from RHIC Run 3, and 13 million central Au+Au collisions from RHIC Run 4 after event cuts. The d+Au and Au+Au data sets were both taken  at $\rts = 200$ GeV. Triggering, centrality selection, and tracking employed
standard STAR procedures
\cite{star_dau,star_pt,star_highpTcoor}. Primary-vertex tracks within $|\eta|<1$ 
were selected for this analysis using standard quality cuts on the
number of hits in the Time Projection Chamber (TPC) and the distance of closest approach to the primary
vertex, which eliminate fake tracks 
and ensure sufficient momentum resolution for
high-\pt{} measurements
\cite{star_pt,star_dau}. The effect of track merging is a negligible effect in the kinematic range used in this analysis.

The central Au+Au events used in this analysis were selected during
data taking based on signals in the Zero Degree Calorimeters
(ZDC) \cite{star_detector} and a cut on the multiplicity in the Central
Trigger Barrel (CTB) to reject peripheral events. The trigger selected
the most central 12\% of the total hadronic cross section, which we
label ``central Au+Au'' in the following.  The d+Au events used in
this analysis were selected using a minimum bias trigger requiring at
least one beam-rapidity neutron in the ZDC in the Au beam direction
(negative pseudorapidity) accepting 95$\pm$3\% of the d+Au hadronic
cross section \cite{star_dau}.\noindent

\section{$\Delta\eta\times\Delta\phi$ di-hadron correlations}
The event-averaged associated hadron distribution, formed using pairs
of charged primary tracks within certain \pt{} intervals for the
trigger and associated particles, is calculated as:
\begin{equation} 
\dsqnwitharg=\frac{1}{N_{trig}}\frac{1}{\epsilon(\phi,\eta,\Delta\phi,\Delta\eta)}\frac{d^2N_{raw}}{d\Delta\phi\, d\Delta\eta},
\label{eq:di-hadronDist}
\end{equation}
\noindent
where \dphino\ and \deta\ are the azimuthal and pseudo-rapidity
separation of the pair, $N_{trig}$ is the number of trigger particles,
and ${d^2N_{raw}}/{d\Delta\phi\,d\Delta\eta}$ is the measured
di-hadron distribution. The factor $1/\epsilon(\phi,\eta,\Delta\phi,\Delta\eta)$
accounts for the reconstruction efficiency of associated tracks,
determined by embedding simulated single tracks into real events, and
for the limited acceptance in $\eta$ and TPC sector boundaries in $\phi$,
determined by event-mixing. Associated particles have $2\ \GeVc <
\ptassno < \pttrigno$ for consistency with previous results \cite{star_highpTcoor}, except
for a new analysis which directly compares correlations for different \pttrig (Section \ref{altbkg}), where  $2 <\ptassno < 4$ GeV/$c$ was used.


Figure \ref{pr} shows distributions of the associated particle yield defined in Eq. \ref{eq:di-hadronDist} for
central Au+Au events with trigger $3 < \pttrigno < 4$ and $4 <
\pttrigno < 6$ GeV/$c$ (upper panels), and for d+Au events with the
same $\pttrig$ selections (lower panels). A near-side peak centered on $(\detano,
\dphino) = (0,0)$ is evident in all \textcolor{black}{panels}, consistent with jet
fragmentation.  In addition, a significant enhancement of
near-side correlated yield is seen at large \deta{}
for central Au+Au events, but not for d+Au events: the ridge.

In this analysis we examine the shape of the near-side associated
yield distribution in detail via projections on the \deta{} and
\dphino-axis. We characterize the shapes of both the ridge and the
jet-like peak, and study the \pt dependence of the ridge and jet-like
yields.

\section{Ridge shape in $\Delta\eta$}
To study the ridge \textcolor{black}{quantitatively}, the di-hadron
distribution is projected onto the \Deta\ axis in intervals of \Dphi:
\begin{eqnarray}
\label{eq:IDphi}
\left.\dNdEta\right|_{a,b}
\equiv\int_{a}^{b}d\Dphi\frac{d^2N}{d\Dphi{d}\Deta};
\end{eqnarray}
\noindent
similarly for projection onto \Dphi:
\begin{eqnarray}
\label{eq:IDeta}
\left.\dNdPhi\right|_{a,b}
\equiv
\int_{|\detano| \in [a,b]}d\Deta\frac{d^2N}{d\Dphi{d}\Deta}
.
\end{eqnarray}
\noindent

%
\begin{figure}[t]
\includegraphics[width=0.49 \textwidth]{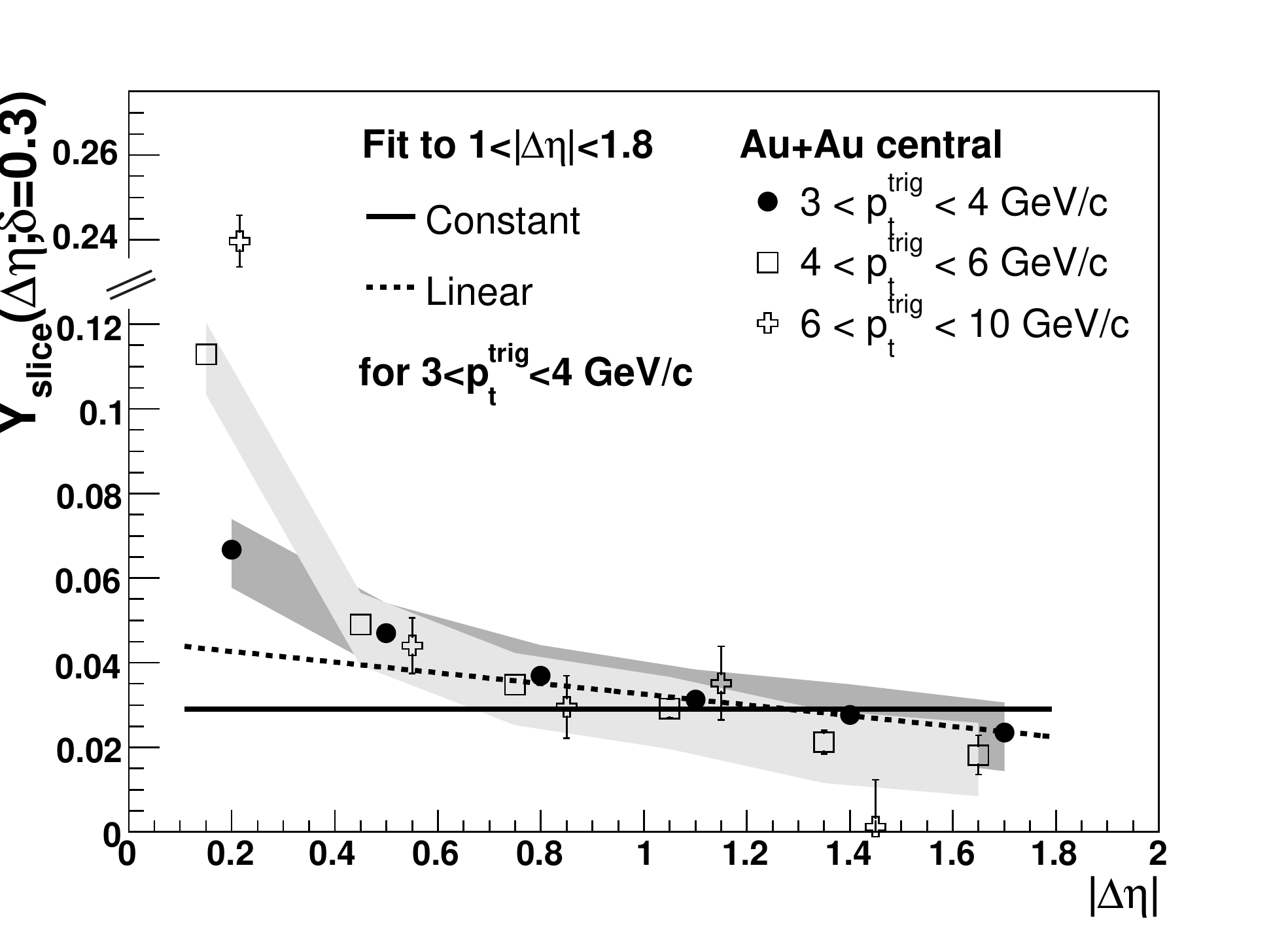}
\caption{\label{fig2}
\YDetadeltathree\ (Eq.\ \ref{eq:Yslice}) for central Au+Au collisions,
$2\ \GeVc < \ptassno < \pttrigno$,  and various
$\pttrigno$ vs. \Deta; \textcolor{black} {the shaded bands represents the
systematic uncertainties due to $v_2$ (not shown for $6 <
\pttrigno < 10$ \GeVc).} The solid and dashed line represents a constant or 
linear fit to $ 1 <  $ \detaabs $ <1.8$; only shown for  $3 < \pttrigno < 4$ \GeVc\ (see text).
Some data points are displaced horizontally for clarity.}
\end{figure}
%

The contribution to the di-hadron distribution of elliptic flow
(\vtwo) in nuclear collisions \cite{star1} is estimated via
\begin{eqnarray}
\label{eq:BDphi}
B_{\Delta\phi}\Argab\equiv b_{\Delta\phi}\int_{a}^{b}d\Dphi
\left(1+2 \langle v_{2}^{trig}v_{2}^{assoc} \rangle \cos2\Dphi\right), 
\end{eqnarray}
\noindent
where the mean uncorrelated level $\bPhi$ is fixed by the assumption
of zero correlated yield at the minimum of the projected distribution,
in this case $1.0 < \Dphi < 1.2$ (zero yield at minimum, or ``ZYAM'')
\cite{star1,star2,zyam,phenixzyam2}. 
Values of $\bPhi$ are given in Table \ref{tab00}.  The modulation
amplitude $\langle v_{2}^{trig}v_{2}^{assoc} \rangle$ is approximated
as $\langle v_{2}^{trig}\rangle \langle v_{2}^{assoc} \rangle$ using
the mean of the event plane, determined at forward rapidities in the Forward TPCs (FTPC), ($v_2\{FTPC\}$(\ptno)) and four-particle
cumulant methods ($v_2\{4\}$(\ptno)) \cite{starflow} (see Table
\ref{tab0}). The $v_2$ systematic uncertainty is defined using
$v_2\{FTPC\}$ as maximum and $v_2\{4\}$ as minimum in each $p_t$
bin. An alternative to the ZYAM procedure is
discussed below (Section \ref{altbkg}).


\begin{table}
\vskip 0.25cm
\begin{center}
\begin{tabular}{cccc}
\pttrig [\GeVc] & $b_{\Delta\phi}$ &  $b_{\Delta\eta}$\\[3pt]
\hline\\[-8pt]
3 $-$ 4 & $0.4302\pm(1\ 10^{-4}$) & $0.1245\pm(1\ 10^{-4}$) \\[2pt] 
4 $-$ 5 & $0.4502\pm(4\ 10^{-4}$) & $0.1296\pm(3\ 10^{-4}$) \\[2pt]
5 $-$ 6  & $0.4533\pm(8\ 10^{-4} $) & $0.1295\pm(6\ 10^{-4}$) \\[2pt]
6 $-$ 10 & $0.4508\pm(1\ 10^{-3}) $ & $0.1284\pm(1\ 10^{-3}$) \\[2pt]
\hline
\end{tabular}
\end{center}
\vskip -0.25cm
\caption{
$b_{\Delta\phi}$  and $b_{\Delta\eta}$ values used
for the ZYAM normalization Eq.\ \ref{eq:BDphi} and \ref{eq:NJeta} for different 
\pttrig windows, with 2 GeV/$c  <
\ptassno < \pttrigno$. Errors are statistical only.
}

\label{tab00}
\end{table}


\begin{table}
\vskip 0.25cm
\begin{center}
\begin{tabular}{ccccc}
\pttrig [\GeVc] & $v_2$(Mean) [\%] &  \ptass [\GeVc]  & $v_2$(Mean) [\%]\\[3pt]
\hline\\[-8pt]

3 $-$ 4 & $8.5\pm 2.2$ & 2.0 $-$ 2.5 & $8.0\pm 1.7$ \\[2pt] 
4 $-$ 5 & $7.7\pm 1.9$ & 2.5 $-$ 3.0 & $8.4\pm 1.9$ \\[2pt]
5 $-$ 6  & $6.5\pm 1.5$ & 3.0 $-$ 3.5 & $8.5\pm 2.1$ \\[2pt]
6 $-$ 10 & $4.6\pm 1.3$ & 3.5 $-$ 4.0 & $8.3\pm 2.1$ \\[2pt]
\hline
\end{tabular}
\end{center}
\vskip -0.25cm
\caption{
Elliptic flow $v_2$ values for different 
\pttrig and \ptass windows, defined as the 
mean of the FTPC reaction plane ($v_2\{FTPC\}$(\ptno)) and four-particle cumulant methods
($v_2\{4\}$(\ptno)) in central Au+Au collisions ($v_2$(Mean)).
Uncertainties are the variation in $v_2$ 
due to these two approaches.}
\label{tab0}
\end{table}


The near-side correlated yield within a \deta interval of width
$\delta$ is then
\begin{eqnarray}
\label{eq:Yslice}
\YDetadelta & = & 
\int_{-0.7}^{0.7}d\Dphi\left(
\left.\dNdPhi\right|_{\Delta\eta-\delta/2,\Delta\eta+\delta/2}
\right) 
\nonumber \\
& - & B_{\Delta\phi}\Argpseven.
\end{eqnarray}
\noindent
The systematic uncertainty on \YDetadelta\ includes contributions from \vtwo,
but not from the ZYAM assumption. The statistical error of $\bPhi$, which is determined independently for every \deta interval using the ZYAM procedure, (and
$\bEta$ Eq.\ \ref{eq:NJeta}) is included in the error on \YDetadelta\ (and \Yridge\ in Eq.\ \ref{eq:Yridge}).

Figure \ref{fig2} shows \YDetadeltathree\ as a function of
\Deta. $Y_{slice}$ is largest around $\deta = 0$, as expected from jet
fragmentation. However, a significant associated yield is also seen at
large $\deta > 1$, for all \pttrigno. The systematic uncertainties in
the figure are due to the uncertainty on the elliptic flow of the
background which may be \detano-dependent. The yield at large \deta\
exhibits no significant dependence on \deta within the experimental
acceptance and the statistical and systematic uncertainties.

A fit to the three data points at largest \detaabs in Fig.~\ref{fig2} 
for the \pttrig intervals 3--4, 4--6 and 6--10
\GeVc, was used to estimate the total ridge yield,
using the two assumptions of no \deta dependence or linear ridge
variation with \detano. These two cases delimit the unknown ridge
yield at small \detano. For all
\pttrigno\ bins the assumption of linear variation gives an estimated
total ridge yield that is 10$-$15\% larger than the assumption of
a \deta-independent ridge. The following discussion assumes that the
ridge is independent on \deta, but the systematic uncertainty assigned
to the ridge yield includes the linear-variation case.

\section{Characterization of jet-like peak} 


\begin{figure}
\includegraphics[width=0.49 \textwidth]{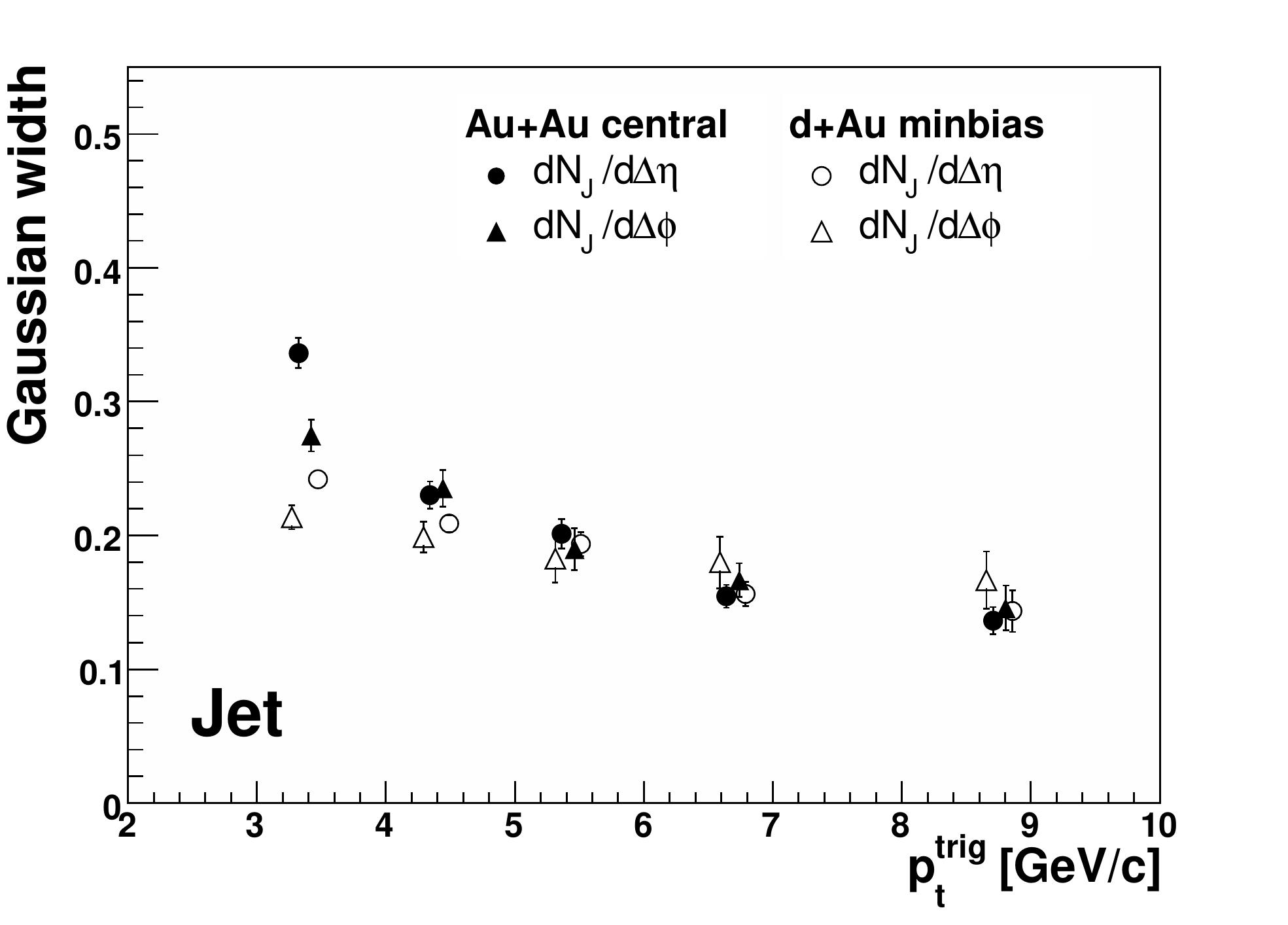}
\includegraphics[width=0.49 \textwidth]{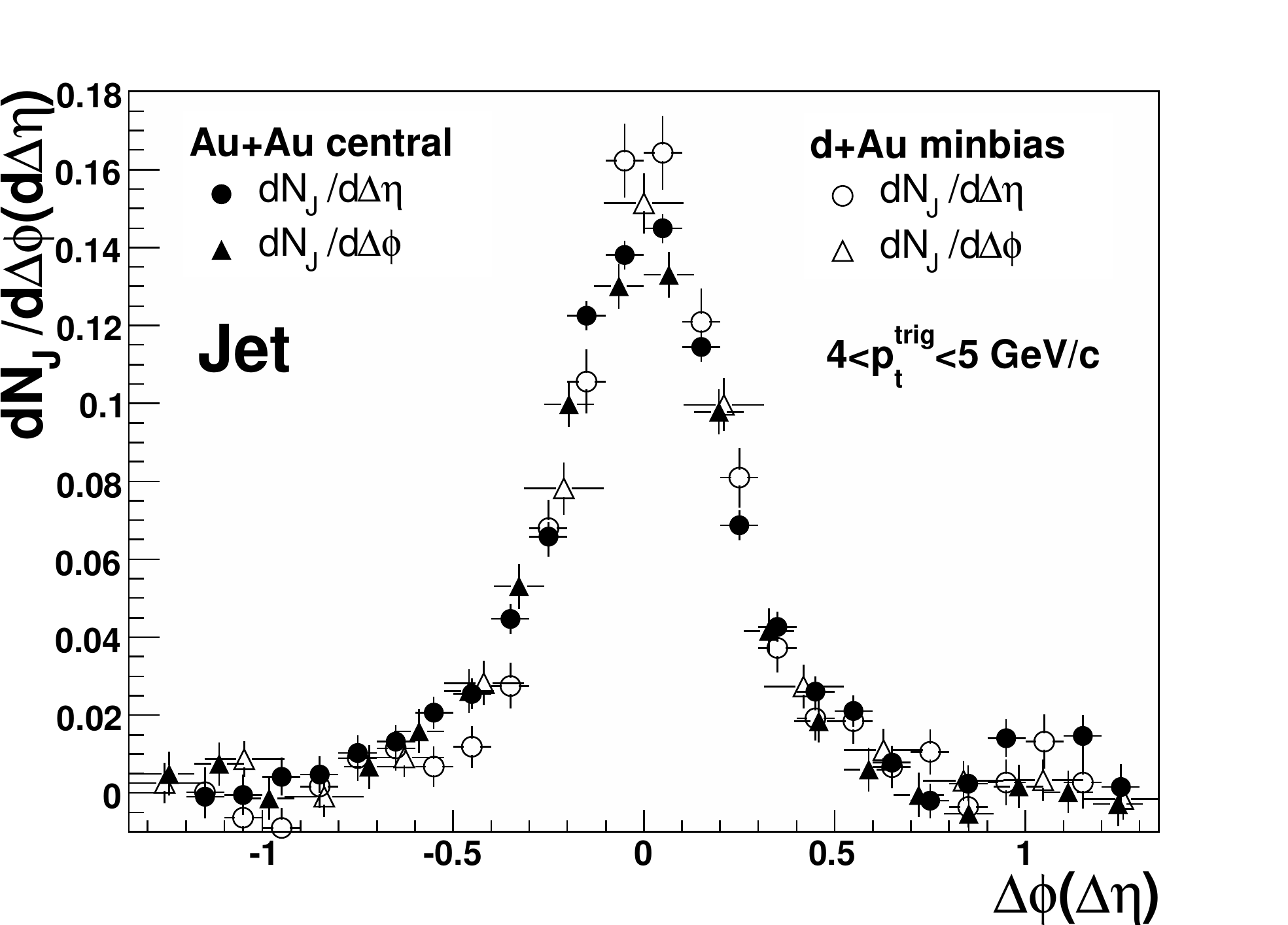}
\caption{\label{fig:widths}
Upper panel: width of Gaussian fit to jet-like peak
for Eq.~\ref{eq:NJeta} (circles) and Eq.~\ref{eq:NJphi} (triangles),
2 GeV/$c  <
\ptassno < \pttrigno$, as a function of
\pttrig, for central Au+Au (filled) and d+Au (open). Some data points are displaced horizontally for clarity.
Bottom panel: the distributions Eq.~\ref{eq:NJeta} and 
Eq.~\ref{eq:NJphi} for $4 < \pttrigno < 5$ \GeVc{} and 2 GeV/$c  <
\ptassno < \pttrigno$.}
\end{figure}


Based on these observations,
we separate the near-side projection onto the \deta\ axis in a jet-like peak
centered at \deta=0 and a \deta-independent ridge
component. Since $v_2$ measured within the acceptance of this analysis has negligible
variation with $\eta$ \cite{starflow}, the jet-like yield may be written as
\begin{eqnarray}
\label{eq:NJeta}
\frac{dN_{J}}{d\Deta}\left(\Deta\right)=
\left.\dNdEta\right|_{-0.7,0.7}-b_{\Delta\eta}
\end{eqnarray}
\noindent
where the constant background level $\bEta$ is calculated in the
interval $1.0<|\Delta\eta|<1.7$. Values of $\bEta$
are given in Table \ref{tab00}.

The jet-like yield can alternatively be defined by projecting onto the
\dphi\ axis, assuming negligible jet-like contribution in $|\Delta\eta|>0.7$:
\begin{eqnarray}
\label{eq:NJphi}
\frac{dN_{J}}{d\Dphi}\left(\Dphi\right)=
\left.\dNdPhi\right|_{0,0.7}-\left.\dNdPhi\right|_{0.7,1.4}
\end{eqnarray}
\noindent
Figure~\ref{fig:widths}, upper panel, shows the widths of Gaussian fits to 
$dN_J/d\deta$ and $dN_J/d\dphi$ vs. \pttrigno, for both central Au+Au
and d+Au events. At low \pttrig the jet-like peak is significantly
broadened in Au+Au relative to d+Au. Similar broadening has been observed previously at
low \pttrig\ \cite{star2,star_deta}.  

For $\pttrigno > 5$ GeV/$c$ the jet-like peak has similar width in
$\Delta\eta$ and $\Delta\phi$ consistent with d+Au reference
measurements.  The full distributions for the two projections are
shown in Fig.~\ref{fig:widths}, lower panel, for central Au+Au and d+Au
minimum bias, for $4<\pttrig<5$ and $\ptass>2\ \GeVc$. The similarity
suggests that for high \pttrigno, the near-side jet-like peak arises
from jet fragmentation in vacuum, with little modification by the
medium for $\ptass > 2\ \GeVc$. \textcolor{black}{Note that this
observation does not preclude significant jet energy loss prior to
fragmentation of the leading parton.}

\section{Characterization of the ridge, \pttrig dependence}


\begin{figure}
\includegraphics[width=0.49 \textwidth]{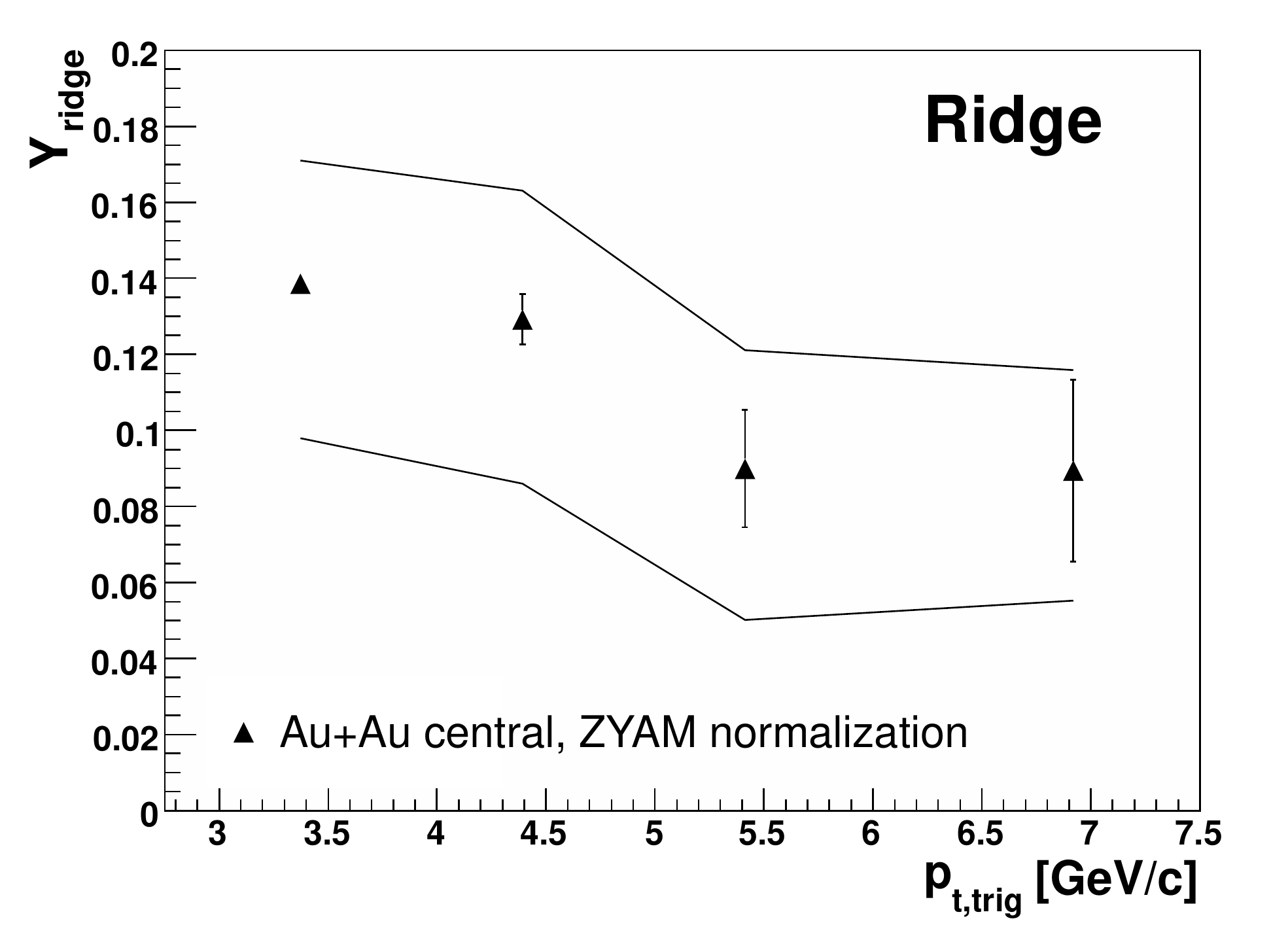}
\caption{\label{fig:yieldpt} Ridge yield (Eq.~\ref{eq:Yridge}) in
$|\detano|<1.7$ and $2\ \GeVc < \ptass < \pttrigno$ as function of
\pttrigno. Solid lines are the systematic uncertainty due to $v_2$. }
\end{figure}


The yield of the jet-like peak $\Yjeta$ and the yield of the ridge $\Yridge$ are 
obtained by suitable integrals over Eqs.\ \ref{eq:Yslice} and \ref{eq:NJeta}:

\begin{align}
\Yjeta =  \int_{-0.7}^{0.7} d\detano \; &
\frac{dN_{J}}{d\Deta}\left(\Deta\right)\,, \label{eq:Yjet} \\
\Yridge =  \int_{-1.7}^{1.7} d\detano \; &
\left.\dNdEta\right|_{-0.7,0.7} \notag \\
& -  B_{\Delta\phi}\Argpseven - \Yjeta. \label{eq:Yridge}
\end{align}
\noindent

Figure \ref{fig:yieldpt} shows \Yridge\ as function of
\pttrig for central Au+Au collisions, using ZYAM to normalize the
background level. Significant ridge yield is observed for all
$\pttrig$, in particular for $\pttrig > 6$ \GeVc, where jet
fragmentation is expected to be the dominant hadron production
mechanism even in nuclear collisions \cite{star_LamK,star_piprot}.

\subsection{Independent estimate of lower bound on ridge yield} 
\label{altbkg}

The above conclusion relies on the two-component model of jet and
background, together with the ZYAM background normalization assumption
and \vtwo\ correction for background.  However, ZYAM does not provide
a strict lower or upper bound on the combinatorial di-hadron
background. An estimate of the combinatorial background that is
systematically independent of these assumptions can be obtained by
attributing the recoil yield entirely to elliptic flow and comparing
near-side yield (small \Dphi ) to the recoil yield in the ridge region
$|\Deta|>0.7$.

Since finite jet-correlated recoil yield 
has been observed over background for $\pttrig>6$ GeV/$c$ and
$\ptass{}>2$ GeV/$c$ \cite{star_highpTcoor}, this procedure will
overestimate the near-side combinatorial background and therefore
underestimate the extracted ridge yield. Full correction for this
effect requires theoretical modelling that is beyond the scope of the
present work. In order to minimize this effect we utilize the
observation that the combinatorial background level is dominantly a
function only of \ptass \cite{star_highpTcoor} and estimate the
maximum background yield for all \pttrig using the recoil distribution
for $4<\pttrig<6$ GeV$/c$, where the jet-correlated recoil yield in
central events is small compared to the background for $\ptass>2$
\GeVc{}. A possible multiplicity bias due to the presence 
of a high-$p_t$ trigger particle is estimated to be around 0.1\%. 
This provides a significant lower bound to the ridge yield.

\begin{figure}[t]
\begin{center}
\includegraphics[width=0.49 \textwidth]{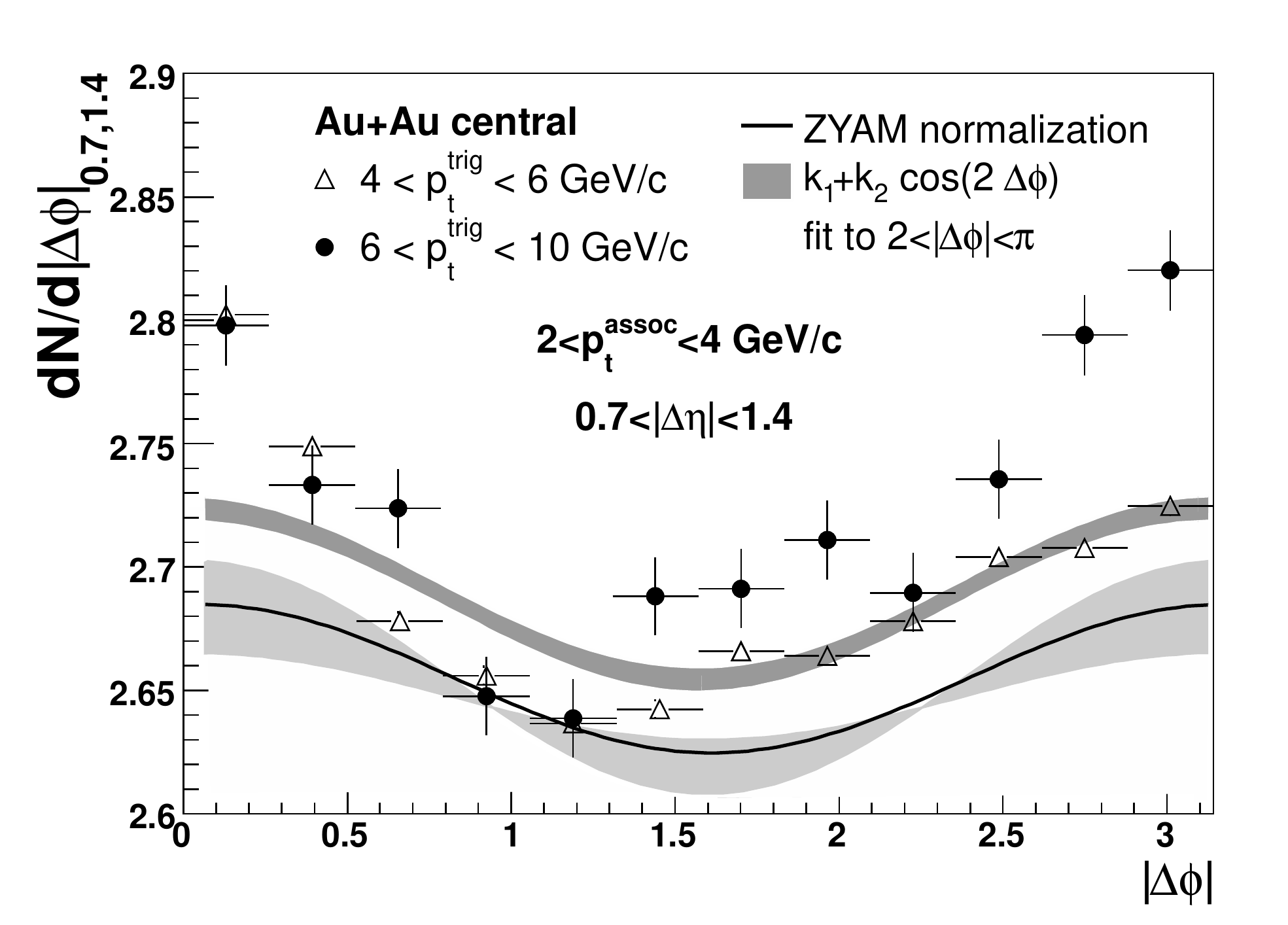}
\caption{\label{recoil}
Projection $dN/d\Delta\phi |_{a,b}$ for 0.7 $<$ \detaabs $<$ 1.4 (Eq.\
\ref{eq:IDeta}) 
in two trigger \pt windows, for $2 < \ptassno < 4$ \GeVc. No background subtraction has been
applied; note the suppressed zero on the vertical scale. The
shaded band shows the fit of the function $k_1+k_2\cdot\cos(2\dphi)$ 
to the recoil region 2 $<$ \dphiabs$<\pi$ for $4<\pttrigno<6$ GeV/c. The width of the band
indicates the fitting error. The solid curve represents the
background estimate using the ZYAM normalization for $4<\pttrigno<6$ GeV/c (systematic
uncertainties are indicated by light shaded band).}
\end{center}
\vskip -0.35cm
\end{figure}

Figure \ref{recoil} shows $dN/d\Delta\phi|_{0.7,1.4}$ (Eq.\
\ref{eq:IDeta}) for $6 < \pttrig < 10$ GeV/$c$ (solid circles) and
$4<\pttrig<6$ GeV/$c$ (open triangles), for $2 < \ptass < 4$ \GeVc. No
background correction has been applied; note the suppressed zero on
the vertical axis. Error bars are statistical. The systematic
uncertainty of associated hadron yields is dominated by a $5\%$ uncertainty
in the tracking efficiency $\epsilon(\Delta\phi,\Delta\eta)$ in
Eq.~\ref{eq:di-hadronDist}. The high-\pt{} tracks used for these
distributions are long, relatively straight tracks with similar
topology, so that the uncorrelated systematic uncertainty of the
distributions in Fig.~\ref{recoil} is negligible relative to the
statistical errors.

The shaded band in the figure shows the fit of the function
$k_1+k_2\cdot\cos(2\dphi)$ to the recoil distribution in the region $2
< \dphiabs < \pi$ for $4<\pttrigno<6$ GeV/$c$. A small, but significant
excess is seen at $|\Dphi|<0.5$ for the signal relative to the band
for both $4<\pttrigno<6$ GeV/$c$ and $6<\pttrigno<10$ GeV/$c$,
corresponding to the ridge yield. The measured distributions
undershoot this background estimate in the region $0.5<\Dphi<1.5$,
which may indicate that this method overestimates the background
somewhat, due to the presence of a small recoil yield even for the
lower \pttrig{} selection. \textcolor{black}{However, it is also
possible that the presence of a trigger locally depletes the
correlated yield relative to an uncorrelated background, in which case
the ZYAM procedure would underestimate the background and the
alternative procedure would be more appropriate.}

The solid line with the light shaded band around it in the figure 
indicates the combinatorial background
estimation using the ZYAM assumption. By construction, this assumption
does not admit an undershoot of the measured distribution relative to
the background. Larger ridge yield is estimated using this technique.

Based on these two independent estimates for the background level and
shape, we conclude that significant near-side ridge yield is present
for $6<\pttrigno<10$ GeV/$c$, indicating that the ridge is indeed
correlated with jet production in central Au+Au collisions.

%
\begin{figure*}[ht]
\includegraphics[width=0.49 \textwidth]{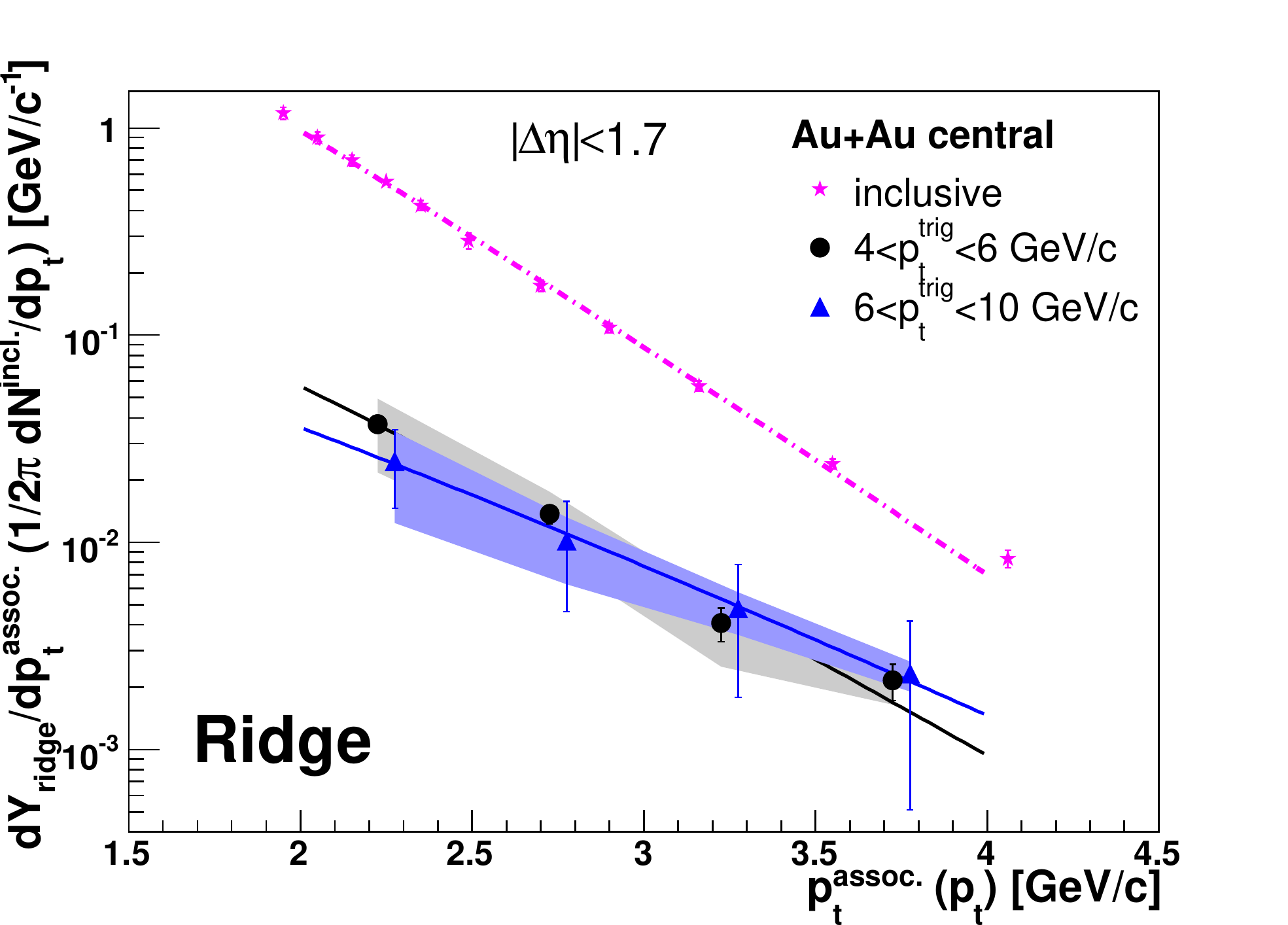}
\includegraphics[width=0.49 \textwidth]{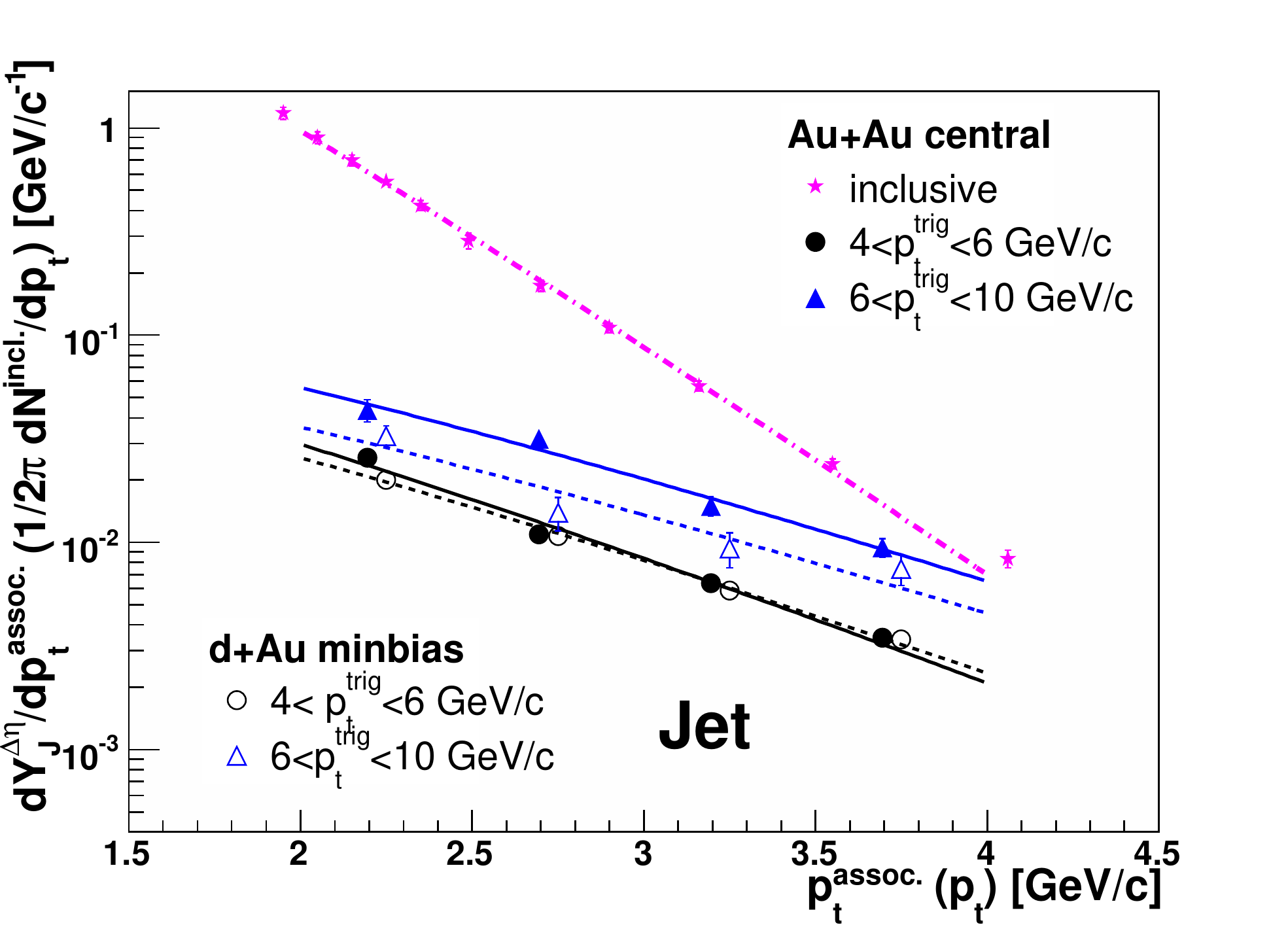}
\caption{\label{fig4} (color online)
Differential \pt\ spectrum for associated particles in
central Au+Au collisions, with $4 < \pttrig < 6$ and $6 < \pttrig < 10$
\GeVc. The dash-dotted line is the inclusive hadron spectrum from central Au+Au collisions \cite{star_pt}.
Left panel: ridge spectrum; shaded bands show systematic
uncertainty. Right panel: jet-like spectrum, also compared to d+Au
reference measurements.  The lines in both panels show exponential
fits to the data (Table~\ref{tab1}). Data are offset horizontally for clarity.}
\vskip -0.35cm 
\end{figure*}


\section{Characterization of the ridge, \ptass dependence}
Figure \ref{fig4} shows the \pt\ spectrum of associated particles in the
ridge, $Y_{ridge}$ (left), and in the jet-like peak, $Y_J^{\deta}$ (right)
(see Eq.~\ref{eq:Yridge} and \ref{eq:Yjet}).  We only consider
$\pttrig > 4$ \GeVc, where the jet-like peak as
defined here is symmetric in \deta\ and \dphi\ and the
peak widths are similar to the d+Au reference
measurements. Table~\ref{tab1} characterizes the spectra through their
inverse slope parameter $T$ from the fit of an exponential function,
$\frac{dN}{dp_t}\propto p_t e^{-p_t/T}$.

The jet-like spectrum is significantly harder than the inclusive spectrum and similar to 
the d+Au
reference measurement, while the ridge-spectrum is softer and more similar to the inclusive spectrum. 
For $4 < \pttrig < 6$ \GeVc, the normalized
jet-like yield per trigger is similar in central Au+Au and d+Au, while at
$6 < \pttrig < 10$ \GeVc{} the yield is slightly
enhanced in Au+Au collisions. 


\begin{table}
\vskip 0.25cm
\begin{center}
\begin{tabular}{cccc}
\pttrig [\GeVc] & $T_{Ridge}$ [MeV/$c$] &  $T_{Jet}$ [MeV/$c$]  & $T_{Jet}^{dAu}$ [MeV/$c$]\\[3pt]
\hline\\[-8pt]
4 $-$ 6 & 416 $\pm$ 22 & 598 $\pm$ 21 & 647 $\pm$ 24\\[2pt] 
6 $-$ 10 & 514 $\pm$ 148 & 702 $\pm$ 47 & 723 $\pm$ 86\\[2pt] 
\hline
\end{tabular}
\end{center}
\caption{Slope parameter $T$ from an exponential fit (see Fig.\ \ref{fig4}) to the \ptass spectrum in different \pttrig bins for ridge-like ($T_{Ridge}$) and jet-like ($T_{Jet}$) near-side correlations as well as $T_{Jet}^{dAu}$ for the d+Au reference measurement (statistical error only). The slope of the inclusive spectrum is $T = 355 \pm 6$ MeV/$c$.} 
\label{tab1}
\vskip -0.35cm
\end{table}


\section{Discussion and Summary}

The similarity of peak shape and \pt distribution of the jet-like
yield in central Au+Au and d+Au collisions, in
contrast to the softer \ptass distribution and the approximately
\detano-independent shape of the ridge yield in
central Au+Au, supports the picture that the near-side correlation at high
\pttrig\ in central Au+Au collisions consists of two distinct
components: a vacuum jet fragmentation contribution, similar to that
seen in p+p and d+Au reference measurements; and the ridge
contribution, with properties similar to bulk particle
production.


Currently available models of ridge formation
\cite{Armesto_flow,Majumder,Romatschke,Wong,Voloshin:2003ud,
Shuryak,Hwa_flow,CGCRidge,CGCRidge2} provide only qualitative guidance
about the underlying physics of the ridge, but not quantitative
predictions at sufficient precision to exclude a given picture based
on the present measurements. All current models generate a softer
spectrum for the ridge yield than for jet-like associated yield, and
describe qualitatively the results in Fig.\ 6. The models involving
turbulent color fields \cite{Majumder,Romatschke} predict a broadening
of the jet-like peak in \detano, which is not observed in these
measurements at high \pt. The observed longitudinal extent of the
ridge ($|\Delta\eta|>1.5$) indicates qualitatively that the ridge is formed early in the evolution of
the fireball, for example as color flux tubes from a CGC initial
state \cite{CGCRidge,CGCRidge2}, and disfavors the gluon radiation
\cite{Armesto_flow} and the turbulent color field mechanisms
\cite{Majumder,Romatschke}, which invoke final state partonic energy
loss and (subsequent) coupling of the radiated gluons to the bulk
matter or color fields. The momentum kick model \cite{Wong} and the
trigger bias model \cite{Voloshin:2003ud, Shuryak} may accommodate the
width of the ridge, though with assumptions about the momentum
distribution and density of the thermal background and the radial flow
boost of the underlying p+p event that are not yet constrained by
available data on inclusive yields and spectra, as well as other
correlation measurements. Another model attributes the ridge structure
to heating of the medium and hadronisation by quark recombination from
QCD matter and seems to reproduce preliminary versions of the
measurement \cite{Hwa_flow}, but does not treat longitudinal dynamics
explicitly. We anticipate that the measurements of the correlation
shapes and yields in this paper will lead to a reassesment of the
various models, and a more quantitative confrontation of the models
with the measurements.

In summary, analysis of di-hadron \detano$\times$\dphi correlations in
central Au+Au collisions reveals a more complex structure of the
near-side correlation than expected from p+p and d+Au reference
measurements, namely the observation of additional correlated yield
at large \deta (the ridge). New detailed measurements of the shape
and the \pttrig and \ptass dependence of the ridge- and jet-like contributions
support the picture that the near-side two-particle correlation
consists of two distinct components: a \detano-independent ridge
contribution with properties similar to inclusive particle production,
and a jet contribution similar to that seen in p+p and d+Au reference
measurements. Various mechanisms have already been proposed for the
formation of the ridge in heavy-ion collisions. The measurements
presented here are expected to rule out or constrain some of the
proposed models.\\

We thank the RHIC Operations Group and RCF at BNL, and the NERSC Center 
at LBNL and the resources provided by the Open Science Grid consortium 
for their support. This work was supported in part by the Offices of NP 
and HEP within the U.S. DOE Office of Science, the U.S. NSF, the Sloan 
Foundation, the DFG cluster of excellence `Origin and Structure of the Universe', 
CNRS/IN2P3, RA, RPL, and EMN of France, STFC and EPSRC of the United Kingdom, FAPESP 
of Brazil, the Russian Ministry of Sci. and Tech., the NNSFC, CAS, MoST, 
and MoE of China, IRP and GA of the Czech Republic, FOM of the 
Netherlands, DAE, DST, and CSIR of the Government of India,
the Polish State Committee for Scientific Research,  and the Korea Sci. 
\& Eng. Foundation.\\

\bibliographystyle{epj}
\bibliography{ref}

\end{document}